\documentclass[superscriptaddress,11pt]{revtex4-1} 

\usepackage{amsmath,amssymb,graphicx}
\usepackage{siunitx}

\begin{document}

\title{A Step-by-step Guide to the Realisation of Advanced Optical Tweezers}

\author{Giuseppe Pesce}
\email{Corresponding author: giuseppe.pesce@fisica.unina.it}
\affiliation{Dipartimento di Fisica, Universit\`a degli studi di Napoli ''Federico II'', Complesso Universitario Monte S. Angelo, Via Cintia 80126, Napoli, Italy}

\author{Giorgio Volpe}
\affiliation{Department of Chemistry, University College London, 20 Gordon Street, London WC1H 0AJ, United Kingdom}

\author{Onofrio M. Marag\'o}
\affiliation{CNR-IPCF, Istituto Processi Chimico-Fisici, V.le F. Stagno D'Alcontres 37, 98158 Messina, Italy}

\author{Philip H. Jones}
\affiliation{Department of Physics and Astronomy, University College London, Gower Street, London WC1E 6BT, United Kingdom}

\author{Sylvain Gigan}
\affiliation{Laboratoire Kastler Brossel, Universit\'e Pierre et Marie Curie, \'Ecole Normale Sup\'erieure, CNRS, College de France, 24 rue Lhomond, 75005 Paris, France}

\author{Antonio Sasso}
\affiliation{Dipartimento di Fisica, Universit\`a degli studi di Napoli "Federico II", Complesso Universitario Monte S. Angelo, Via Cintia 80126, Napoli, Italy}

\author{Giovanni Volpe}
\affiliation{Soft Matter Lab, Department of Physics, Bilkent University, Cankaya, 06800 Ankara, Turkey}
\affiliation{UNAM-National Nanotechnology Research Center, Bilkent University, Cankaya, 06800 Ankara, Turkey}

\begin{abstract}
Since the pioneering work  of Arthur Ashkin, optical tweezers have become an indispensable tool for contactless manipulation of micro- and nanoparticles. Nowadays optical tweezers are employed in a myriad of applications demonstrating the importance of these tools. While the basic principle of optical tweezers is the use of a strongly focused laser beam to trap and manipulate particles, ever more complex experimental set-ups are required in order to perform novel and challenging experiments. With this article, we provide a detailed step-by-step guide for the construction of advanced optical manipulation systems. First, we explain how to build a single-beam optical tweezers on a home-made microscope and how to calibrate it. Improving on this design, we realize a holographic optical tweezers, which can manipulate independently multiple particles and generate more sophisticated wavefronts such as Laguerre-Gaussian beams. Finally, we explain how to implement a speckle optical tweezers, which permit one to employ random speckle light fields for deterministic optical manipulation.
\end{abstract}

\maketitle 

\section{Introduction}

The study of the interaction between light and matter has been extremely prolific, leading to the invention of novel techniques that have had a profound impact in many different fields. In particular, the use of light to exert forces and torques on microscopic and nanoscopic particles has blossomed: since the pioneering work of Arthur Ashkin \cite{AshkinPRL70,AshkinOL86}, optical tweezers (OT) have become an indispensable tool in, e.g., biophysics, experimental statistical physics and nanotechnology \cite{GrierN03,NeumanRSI04,JonasE08,MaragoNN13}. At the same time, starting from the relatively simple tightly focused laser beam employed by Ashkin and co-workers to confine a microscopic particle in three dimensions \cite{AshkinOL86}, experimental set-ups for optical trapping and manipulation have grown ever more complex, leading to OT capable, e.g., of trapping multiple particles \cite{GrierN03}, of rotating particles by transferring spin  \cite{,BishopPRL04,DonatoNC14} and orbital angular momentum \cite{GrierN03,PadgettNP11}, of measuring forces in the femtonewton range \cite{RohrbachPRL05,ImparatoPRE07}, and of sorting particles based on their physical properties \cite{MacDonaldN03,JonasE08,VolpeSR14}.

Many OT set-ups are built on commercial microscopes, especially for experiments on biological samples such as DNA, molecular motors, and cells \cite{BustamanteN03}. While commercial microscopes offer several advantages, e.g.,  high quality imaging of the sample, easy integration with powerful imaging techniques such as phase contrast, fluorescence excitation and differential interference contrast, they provide limited flexibility when dealing with more elaborate experiments. Thus, many research groups have started to develop home-made microscopes, which have the great advantages of being very versatile, much less expensive, and, if implemented properly, with a superior mechanical stability. 

In the literature, there are various articles that provide some guidance on how to implement an OT set-up. For example, Refs.~\cite{BechhoeferAJP02,AppleyardAJP07} explain how to build and calibrate a single-beam OT set-up for undergraduates laboratories and Refs.~\cite{LeeNP07,MathewRSI09} explain how to transform a commercial microscope into an OT. However, to the best of our knowledge, a detailed explanation of how to build a home-made OT set-up with advanced functionalities such as multiple holographic optical traps and speckle OT is still missing.

In this article, we provide a detailed step-by-step guide to the implementation of state-of-the-art research-grade OT. The only requirement to be able to follow the procedure we explain is a basic knowledge of experimental optics. We start by implementing a high-stability home-made optical microscope and, then, we transform it into a series of OT with increasingly more advanced capabilities. In particular, we explain in detail the implementation of three OT set-ups with cutting-edge performance: (1) a high-stability single-beam OT capable of measuring forces down to few fN \cite{ImparatoPRE07}; (2) a holographic optical tweezers (HOT) \cite{PadgettLOC11} capable of creating multiple trap and advanced optical beams such as Laguerre-Gaussian beams; and (3) a speckle OT particularly suited to develop robust applications in highly scattering environments and in microfluidic devices \cite{VolpeOE14,VolpeSR14}. For each set-up, we also provide a time-lapse movie (see supplementary information) in order to make it easier to understand and follow each step of the implementation.

In Section~II, we provide a brief overview of the set-ups. In Section~III, we provide a detailed step-by-step procedure for the realization of each set-up: first, we build a home-made microscope capable of performing digital video microscopy (DVM) by analysing the images acquired with a CCD camera; then we show how to upgrade this microscope to a single-beam OT equipped with a position sensitive detection device based on a quadrant photodiode (QPD); finally, we describe how to build a HOT and a speckle OT. In Appendix~A, we discuss how to calibrate the single-beam OT. In Appendix~B, we explain the fundamentals of how spatial light modulators (SLM) work and we provide some examples of the most commonly used phase masks and algorithms employed with HOT.

\section{Optical Tweezers set-ups}

\subsection{Home-made microscope}

In the implementation of every OT experiment, it is crucial to start with a microscope that is as mechanically stable as possible. For this reason our home-made microscope (Fig.~\ref{fig:singleot}a and \ref{fig:singleot}b) is built on a stabilized optical table, which permits us to minimize environmental vibrations, and its structure is realized with AISI 316 stainless steel, which offers one of the lowest thermal expansion coefficient and a great rigidity. This structure consists of three levels. The first level, which lies on the optical table, is where the optical components necessary to focus the sample image on the camera (and to prepare the optical beam for the OT, Subsection~II.B) are hosted. In particular, there is a mirror arranged at \ang{45} degrees to reflect the illumination light toward the CCD camera (and the laser beam toward the objective lens for the OT, Subsection~II.B). Videos of the sample are recorded by a fast CCD camera (Mikrotron, MotionBLITZ EoSens Mini 1).

The second level is realized with a breadboard held on eight columns; here is where the objective mount as well as the stages to hold and manipulate the sample are hosted. The objective holder is mounted vertically and comprises a high stability translational stage (Physik Instrumente, M-105.10) with manual micrometer actuator (resolution $1\,{\rm \mu m}$); if needed, this stage can be easily equipped with a piezo actuator for computer-controllable and more precise movements. The sample stage is realized by a combination of a manual stage for coarse (micrometric) movement (Newport, M406 with HR-13 actuators) and a piezoelectric stage (Physik Instrumente, PI-517.3CL) with nanometer resolution.

Finally, the third level, which is build on the second-level breadboard using four columns, hosts the illumination lamp (and the position detector devices, for the photonic force microscope (PFM), Subsection~II.C). To illuminate the sample we opted for a cold-light white LED (Thorlabs, MCWHL5), which offers high intensity without heating the sample. The white LED  is collimated by an aspheric lens. The whole system is easily realized using modular opto-mechanics components (Thorlabs, cage system). As condenser lens we use an objective with magnification $10\times$ or $20\times$ mounted on a five-axis stage (Thorlabs, K5X1), which assures fine positioning of the lens and good stability at the same time.

\subsection{Single-beam optical tweezers (OT)}

The single-beam OT (Fig.~\ref{fig:singleot}a and \ref{fig:singleot}b) is realized by coupling a laser beam to the home-made microscope described in Subsection 2.1. We use a solid state laser featuring a monolithic Nd:YAG crystal NPRO (Non-Planar Ring Oscillator) (Innolight, Mephisto 500, $\lambda = 1064\,{\rm nm}$, $500\,{\rm mW}$ max power), which exhibits excellent optical properties for optical manipulation experiments that require high stability (e.g., for the measurement of femtonewton forces), i.e., extremely low intensity noise  and very good pointing stability. The output of this kind of laser is slightly diverging and features elliptical polarization. Therefore, we linearize its polarization with a zero-order quarter-wave plate. To change the power without altering the injection current of the pump diode laser, which could affect the laser beam emission quality and, therefore, the laser performance, we used the combination of a zero-order half-wave plate to rotate the linear polarization and a linear polarizer to finely tune the laser power. The laser beam is directed through the objective making use of a series of mirrors. A $5\times$ telescope is added along the optical train in order to create a beam with an appropriate size to overfill the objective back aperture and to generate a strong optical trap, as typically the beam waist should be comparable to the objective back aperture in order to obtain an optimal trapping performance \cite{JonasE08}.

In order to focus the laser beam inside the sample cell and generate the optical trap, we use a water-immersion objective (Olympus, UPLSAPO60XW) because water-immersion objectives present less spherical aberrations than oil-immersion objectives, and, thus, a better trapping performance. Ref.~\cite{AlexeevEJP12} provides a detailed comparison of the performance of various objectives.

In order to achieve the highest mechanical stability, the OT set-up is realized keeping the optical train  as short as possible. Nevertheless, the optical train can be modified according to the specific requirements of each experiment. For example, if a movable trap is needed, a beam steering system can be added according to the procedure described in Ref.~\cite{FallmanAO97}.

\subsection{Photonic force microscope (PFM)}

An optically trapped particle scatters the trapping s so that the light field in the forward direction is the superposition of the incoming and scattered light. This basic observation has been harnessed in the development of interferometric position detection techniques \cite{GittesOL98}. In our set-up (Figs.~\ref{fig:qpd}a and \ref{fig:qpd}b), a condenser (Olympus PlanC N, $10\times$) collects the interference pattern arising from the interference between the incoming and scattered fields and a photodetector located on the condenser back-focal plane (BFP) records the resulting signals. As detector, we use a QPD based on an InGaAs junction (Hamamatsu, G6849), which, differently from silicon-based photodiodes \cite{SorensenJAP03}, has a good high-frequency response in the infrared region of the spectrum. The output signals are acquired with a home-made analog circuit \cite{FinerBJ96} and then analyzed in order to obtain the three-dimensional particle position. These signals can then be used to calibrate the optical trap as explained in Appendix~A.

\subsection{Holographic optical tweezers (HOT)}

The experimental set-up for HOT is slightly more complex than the one for a single-beam OT. HOT use a diffractive optical element (DOE) to split a single collimated laser beam into several separate beams, each of which can be focused inside the sample to generate an OT \cite{GrierN03}. These optical traps can be made dynamic and displaced in three dimensions by projecting a sequence of computer-generated holograms. Furthermore, non-Gaussian beam profiles can be straightforwardly encoded in the holographic mask. 

In our implementation (Figs.~\ref{fig:hot}a and \ref{fig:hot}b), the DOE is provided by a spatial light modulator (SLM) (Hamamatsu, LCOS-SLM, X10468-03). The SLM permits us to modulate the phase of the incoming beam wavefront by up to $2\pi$. In practice, the phase mask that alters the beam phase profile is a $8\,{\rm bit}$ greyscale image generated and projected on the SLM using a computer (Appendix~2). We placed the SLM after a $6\times$ telescope, which increases the beam size so that it completely fills the active area of the SLM. Then, we employ two lenses (with equal focal length, $f=750\,{\rm mm}$) to image the SLM plane onto the BFP of the objective (and the necessary mirror to direct the optical train). In practice, the first lens is placed after the SLM at a distance equal to $f$, the second lens is placed at a distance equal to $2f$ from the first lens, and the BFP is at a distance $f$ from the second lens. Thus, the total distance from the SLM to the BFP is $4f$ (for this reason this is often referred to as a $4f$-configuration). We note that it is also possible to use lenses with different focal lengths to change the total path length and magnify/de-magnify the beam, for which we refer the readers to Ref.~\cite{FallmanAO97}. Therefore, the SLM (just like the BFP) is placed at the Fourier plane of the front-focal plane (FFP) of the objective and the field distribution in the FFP is (in first approximation) equal to the Fourier transform of the field distribution at the SLM \cite{PadgettLOC11}.

We built our HOT on the same home-made microscope used for the single-beam OT. Apart from the addition of the SLM and necessary optics in the optical train, the only modification concerns the illumination system: since the QPD is no longer needed, we preferred to increase the illumination intensity at the sample by moving the LED nearer to the condenser. High illumination intensity is needed in order to achieve high acquisition rates in digital video microscopy, as increasing the acquisition rate entails a decrease of the shutter time. For example, when we acquire images ($128\times 100$ pixels) at $18000$ frames per second, the maximum shutter time is only $55\,{\rm \mu s}$.

\subsection{Speckle optical tweezers}

The experimental set-up for speckle OT is a simplified version of the set-up for a single-beam OT. The main difference is that the condenser and illumination are substituted by a multimode optical fiber (core diameter $105\,{\rm \mu m}$, numerical aperture ${\rm NA}=0.22$, Thorlabs, M15L01) that delivers both the trapping laser light and the illumination light directly in close proximity to the sample upper surface, while the trapping beam must be spatially coherent and monochromatic to generate the speckle pattern through interference, the illumination should be generated with a spatially and/or temporally incoherent source in order to ensure a uniform illumination of the sample. The speckle light pattern is generated by directly coupling the laser beam into the multimode optical fiber \cite{MoskNP12}. The random appearance of the speckle light patterns (Fig.~\ref{fig:speckle}d) is the result of the interference of a large number of optical waves with random phases, corresponding to different eigenmodes of the fiber. More generally, speckle patterns can be generated by different processes: scattering of a laser on a rough surface, multiple scattering in an optically complex medium, or, like in this work, mode-mixing in a multimode fiber  \cite{MoskNP12}. The choice of a multimode optical fiber provides some practical advantages over other methods, namely generation of homogeneous speckle fields over controllable areas, flexibility and portability in the implementation of the device, as well as higher transmission efficiency. 

In our set-up (Figs.~\ref{fig:speckle}a and \ref{fig:speckle}b), the fiber output is brought in close proximity of the upper wall of the sample by a micrometric three-axis mechanical stage. Optical scattering forces push the particles in the direction of light propagation towards the bottom surface of the microfluidic channel, so that they effectively confine the particles in a quasi two-dimensional space. The particles are then tracked by digital video microscopy. 

\section{Step-by-step procedure}

\subsection{Home-made inverted microscope}

\begin{enumerate}
\itemsep1em

\item Build the first two levels of the microscope structure on a vibration-insulated optical table. The first level can be either the optical table itself or, as in our case, a custom designed breadboard. The main advantage of using an independent breadboard as first level is that the microscope is completely stand-alone and can be easily moved, if necessary. The optical components necessary to align the optical beam and to image the sample on the camera will be hosted on the optical table. Particular care should be taken in positioning the microscope: it is preferable to mount it at the center of the optical table, since placing it on a side could decrease the damping performance of the table. The second level is  composed of a breadboard held on four columns; here is where the stages to hold and manipulate the sample will be hosted. Verify that these planes are stable and horizontal using a bubble-level. (The third level will be mounted later and will host the illumination white LED, step 5, and the quadrant photodiode for particle tracking, Subsection~III.C.)

\item Fix a \ang{45} gold mirror (M1, Fig.~\ref{fig:singleot}a and \ref{fig:singleot}b) on the first level of the structure. This mirror will serve to deflect the white light from the microscope towards the camera (and to deflect the laser beam towards the back aperture of the objective, Subsection~III.B). The mirror center should be aligned with the central vertical axis of the structure (which will correspond to the axis of the objective). The height of the white light beam (and of the laser beam) above the optical table is determined by the height of the center of the \ang{45} mirror. In our particular case, it is  fixed at $7\,{\rm cm}$. 

\item Fix on the bottom side of the second level breadboard the vertical translation stage where the microscope objective will be mounted. The vertical stage has to be perfectly perpendicular to the breadboard. Use a right angle bracket to check the alignment (thanks to the length of the objective holder, angles as small as a few tenths of degree can be easily detected). Since the vertical stage should be firmly attached to the second level, it is preferable to avoid kinematic mounts even if of the best quality. Instead, to correct small misalignments we suggest putting thin micrometric spacers between the mechanical parts and then tightly fastening the screws. This technique will lead to a very precise vertical alignment of the objective with respect to the vertical axis of the microscope. 

\item Mount the sample translation stage on the upper side of the second-level breadboard. In general, this translation stage can be either manual or automatized. Manual translation stages typically offer precision down to a fraction of a micrometer over a range that can easily reach several millimeters. Automatized translation stages, which are often driven by mechanical or piezoelectric actuators, permit one to achieve subnanometer position precision over a range that typically does not exceed a few tens or hundreds of micrometers and have the advantage that they can be controlled in remote. In the realization of our set-up, we combine a manual $xy$-translation (horizontal) stage for the rough positioning of the sample and an automatized piezoelectric $xyz$-translation stage for fine adjustment.

\item Proceed to build the third level of the structure, which is constituted of a breadboard held by four columns and will host the illumination and detection components.

\item Place a five-axis translation stage on the bottom side of the third level plane to mount the condenser. This stage permits one to align the condenser position along the lateral (horizontal) directions and, therefore, to center the position of the collected illumination beam. The alignment along the vertical direction will be performed at step 12 to optimize the illumination of the sample.

\item Mount the objective to collect the image of the sample (and later to focus a laser beam for the OT, Subsection~III.B). At the beginning, it is convenient to use a low-magnification (for example, $10\times$ or $20\times$) objective to simplify the alignment. Later, a high-numerical aperture water- or oil-immersion objective (in our case, a water-immersion objective, Olympus, UPLSAPO60XW) will be needed for optical trapping. Using objective lenses with the same parfocal length guarantees that switching between them will not modify the relative position of the image plane with respect to the objective stage (and, thus, it will not require any adjustment of the objective position).

\item Mount the condenser. This lens can be either a $10\times$ or $20\times$ infinity corrected objective or an aspheric condenser lens. The choice depends on the position detection system: the aspheric condenser lens is preferable for digital video microscopy because it permits higher and more uniform illumination intensity; the objective lens is preferable if the detection is based on a QPD because it permits one to collect the scattered light over a larger angle (Subsection~III.C). The condenser position along the vertical axis will be optimized later (step 12).

\item Place a sample on the sample stage. Start by using a calibration glass slide (best choice to set up the microscope), or a sample of microspheres stuck to a coverslip, or a glass coverslip to which some color has been applied on one side, e.g., by using an indelible marker. 

\item Add an illumination source on the third level to illuminate and observe the sample. A white LED lamp is better than a halogen lamp because it will not heat the sample. The light from the LED can be collimated using an aspheric lens. The collimated light is then focused on the sample by the condenser. It is possible to see the resulting image by the naked eye by placing a white screen along the path of the light collected by the objective. The image of a calibration glass slide can be easily placed in the centre of the circular spot of the light transmitted by the objective lens. With the illumination at the maximum intensity, project this image at the longest possible distance ($>5\,{\rm m}$). Adjust the focus position to have a clear image of the sample. This will set the objective in the correct position to image the sample to infinity, i.e., the objective is properly imaging objects that are in  the image plane. Once this procedure is done, do not move the relative position between the objective and the sample until the CCD camera is placed in its final position (next step).

\item Add a CCD camera. When using infinite-corrected objectives, it is necessary to use a tube lens (L1) to produce the image of the sample at a specific position. The CCD sensor should be placed at a distance from the tube lens equal to its focal length. Place the tube lens (L1) and then the camera. If needed, it is possible to use a some mirrors (M2 and M3) to place the lens and the camera in a more convenient position.  The alignment of the camera and of the tube lens is crucial to have a clear image and to minimize distortions; in particular: (1) the white light spot must be in the center of the lens and the lens face must be perpendicular to the light path; (2) the heights of the tube lens and of the camera must be equal; (3) the mirrors along the image path should be adjusted so that the light beam makes \ang{90} angles (the screw holes of the optical table may be useful as guides in the alignment procedure). The focal length of the tube lens determines the magnification of the microscope with a fixed objective: the longer the focal length, the higher the magnification. Nevertheless, increasing the focal length reduces the amount of light that reaches the camera and, thus, limits the acquisition speed. Therefore, a compromise between magnification and acquisition speed should be reached. To avoid damaging the camera sensor, perform this whole procedure with low lamp power, well below the camera saturation threshold.

\item Optimize the  illumination brightness by adjusting the condenser axial position. More complex illumination configurations, e.g., K\"ohler illumination, are also possible if the coherence of the illumination as well its homogeneity need to be controlled. 

\item It is useful to calibrate the CCD measuring the pixel-to-meter conversion factor. This can be done using a calibration glass slide or moving (by a known controllable displacement) a particle stuck to the bottom coverslip by means of a calibrated piezoelectric stage.

\end{enumerate}

\subsection{Single-beam OT and alignment of the laser beam}

\begin{figure}[ht]
\centerline{\includegraphics[width=0.7\columnwidth ]{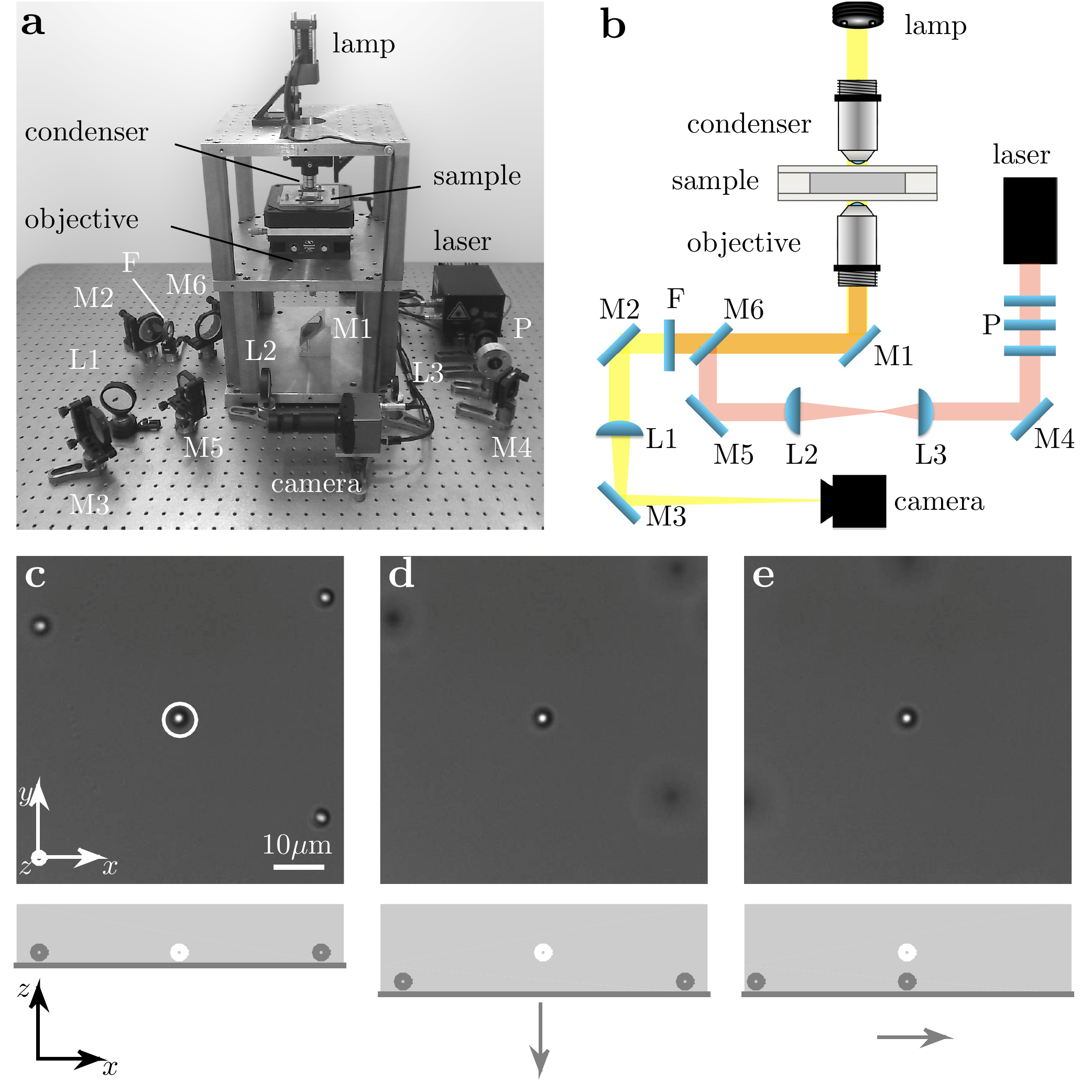}}
\caption{Single-beam optical tweezers (OT). (a) Photograph and (b) schematic of the experimental set-up. See Movie 1 for the complete building sequence. (c) A $1\,{\rm \mu m}$ diameter polystyrene bead (circled particle) is trapped by the OT and then the stage is moved (d) vertically and (e) laterally; notice how, as the stage is moved, the trapped particle remains in focus in the image plane, while the background particles get defocused.}
\label{fig:singleot}
\end{figure}

\begin{enumerate}
\itemsep1em
\addtocounter{enumi}{13}

\item Fix firmly on the optical table a laser. In our case, we use a solid state laser with wavelength $1064\,{\rm nm}$ and maximum power $500\,{\rm mW}$, even though typically we will only need $1\,{\rm mW}$ for trapping a single particle. Place the laser so that the beam height coincides to the height of the center of the \ang{45} mirror (M1, Fig.~\ref{fig:singleot}a and \ref{fig:singleot}b) on the first level of the microscope (step 2). This is the height of the axis of the optical train ($7\,{\rm cm}$ in our case). Importantly, keep in mind that correct safety procedures when working with lasers are of paramount importance; thus, we urge all users to undergo suitable safety training before starting to work with lasers and to observe all local safety regulations, including, in particular, the use of appropriate eye safety equipment.

\item Since we use a diode-pumped solid state (DPSS) laser, whose output beam has elliptical polarization, we place a quarter-wave plate to linearize the polarization just after the laser (the first element in P, Fig.~\ref{fig:singleot}a and \ref{fig:singleot}b). In order to maintain the power stability of the laser beam, we prefer not to alter the laser settings (and in particular its injection current) in order to change the laser output power. Therefore, we opt for an alternative way of controlling the laser power: we place along the beam path, a half-wave plate and a beam polarizer (second and third elements in P, Fig.~\ref{fig:singleot}a and \ref{fig:singleot}b), which permit us to tune the beam power by rotating the half-wave plate. The next step is to mount the optical train to direct the laser beam through the objective. In order to have a stable optical trap, keep the laser beam path as short as possible and, also, consider using black plastic pipes to enclose the laser beam and to built a thermal and acoustic isolation enclosure for the whole experimental set-up. For safety reason and to avoid damaging the camera, perform this procedure with low laser power.

\item Use dielectric laser line mirrors to deflect the laser beam at right angle (M4 and M5, Fig.~\ref{fig:singleot}a and \ref{fig:singleot}b). Check that the deflected beam travels at the same height of the incident beam. The last mirror of the optical train should be a dichroic mirror (M6, Fig.~\ref{fig:singleot}a and \ref{fig:singleot}b), which permits one to split the light of the laser beam, which is routed towards the objective, and the light of the lamp, which is routed towards the camera. We used a short-pass dichroic mirror that reflects (reflection coefficient $>98\%$) in the region from $750$ to $1100\,{\rm nm}$ and transmits shorter wavelengths.

\item Remove the objective and the condenser from their respective holders and, using only the last two mirrors (M5 and M6, Fig.~\ref{fig:singleot}a and \ref{fig:singleot}b), align the laser beam so that it goes straight through the objective holder. This is a crucial point as the  laser beam must be perfectly aligned along the vertical axis of the objective holder. To achieve this, you can use two alignment tools like pinholes (Thorlabs, VRC4D1) or fluorescing disks (Thorlabs, VRC2RMS). It is not critical to perfectly center the condenser at this point, because it will be optimized later while aligning the detection components.

\item Add a telescope in order to create a beam with an appropriate size to overfill the objective back aperture and to generate a strong optical trap. The beam waist can be measured by fitting a Gaussian profile to an image of the beam acquired by a digital camera (some neutral density filters might be necessary to prevent the camera from saturating). The typical beam waist (radius) of a DPSS laser is about $1$ to $1.5\,{\rm mm}$, therefore a telescope with magnification of about $5\times$ guarantees the proper overfilling condition. The magnification can be easily obtained with lenses with focal lengths of $5\,{\rm cm}$ and $25\,{\rm cm}$. To realize the telescope, first place the lens with the longer focal length (L2, Fig.~\ref{fig:singleot}a and \ref{fig:singleot}b). It should be positioned while controlling the position of the laser beam on the target placed on the objective lens mount: the position of the spot center must remain the same with and without the lens. Then place the lens with shorter focal length (L3, Fig.~\ref{fig:singleot}a and \ref{fig:singleot}b), between the first lens and mirror M4, controlling, at the same time, that the position of the laser beam on the target does not change, as done for the first lens, and the collimation of the laser beam. The latter can be done using a shear plate interferometer or checking that the size of the beam waist is constant on a long travel distance  ($>5\,{\rm m}$).

\item Double-check the alignment of the laser beam along the vertical axis of the microscope using the alignment tools.

\item Add a short-pass filter (Thorlabs, FGS900) in the light path leading to the camera in order to remove any laser light that might saturate the camera (F, Fig.~\ref{fig:singleot}a and \ref{fig:singleot}b). It is preferable to mount the filter on a repositionable mount (Thorlabs, FM90) to easily block or unblock the laser light.

\item Mount again the objective, the condenser and the sample cell. Allow the laser light to reach the CCD camera by removing the filter F (Fig.~\ref{fig:singleot}a and \ref{fig:singleot}b).

\item Place a droplet of the immersion medium on the objective (water in our set-up) and gently move the objective towards the sample cell until the focus is at the glass-air interface. This is best done by using a sample cell without any liquid in order to increase the amount of light reflected back to the camera, but can also be done with a sample filled with a solution, even though the intensity of the back-scattered light will be significantly (about two orders of magnitude) lower. In practice, one has to slowly approach the objective to the sample, until the back-scattered pattern size is minimized. The resulting pattern (a series of concentric Airy rings, possibly with a cross-shaped dark area) should be as symmetric as possible. Furthermore, the pattern should remain symmetric and with the same center as the relative distance between focus and interface is changed. 

\item Build a sample cell with a microscope slide and a coverslip (thickness \#1, $150\,{\rm \mu m}$ ca.): (1) clean the slide and coverslip; (2) place two stripes of Parafilm\textsuperscript{\textregistered} above the slide to work as spacers (Parafilm\textsuperscript{\textregistered} is a polymer resistant to many solvents and acids; it melts at about $80\textsuperscript{$\circ$}{\rm C}$ and it works like a glue when cooled); (3) place the coverslip above the Parafilm\textsuperscript{\textregistered} stripes; (4) heat with a hot air gun or placing the cell on a hot plate; (5) fill the cell with the sample solution and seal it with silicon grease, epoxy resin or UV glue. The thickness of the cell is around 120 $\mu m$, but it is possible to have thicker cells overlapping two or three layers of Parafilm\textsuperscript{\textregistered} or thinner by stretching Parafilm\textsuperscript{\textregistered} before placing the coverslip.

\item Place the sample cell filled with a mixture of water and microspheres on the stage (with the coverslip side below towards the objective) and adjust the position of the objective and sample holder such that the plane just above the glass-solution interface is on focus. The image recorded by the camera should show the sedimented particles, as in Fig.~\ref{fig:singleot}c. 

\item Move the sample stage a few micrometers down (so that the laser focus effectively moves a few micrometers above the interface) and approach some of the particles. If a particle is free to move and not stuck to the glass, you will see the particle jump into the laser beam and get optically trapped. To get the image in focus, it might be again necessary to slightly adjust the lens in front of the camera. It could happen that a bead is attracted by the laser beam but instead  of being trapped it will be pushed in the direction of the laser beam propagation. This means that the laser beam is not well aligned or is not overfilling the aperture sufficiently. Go back to steps~17-19 to check the alignment and overfilling. In order to double-check that the particle is actually optically trapped, try and move it around. First move the sample stage further down so that the optically trapped particle gets moved vertically above the glass-solution interface, as in Fig.~\ref{fig:singleot}d: non-optically trapped particles on the sample cell bottom get out-of-focus, while the optically trapped particle remains in focus. Then, move the sample stage horizontally, as in Fig.~\ref{fig:singleot}e: the image of the optically trapped particle remains on the same spot within the image captured by the camera, while the background particles sedimented on the coverslip appear to be displaced.

\end{enumerate}

\subsection{Photonic force microscope (PFM) and alignment of the quadrant photodetector (QPD)}

\begin{figure}[ht]
\centerline{\includegraphics[width=0.7\columnwidth ]{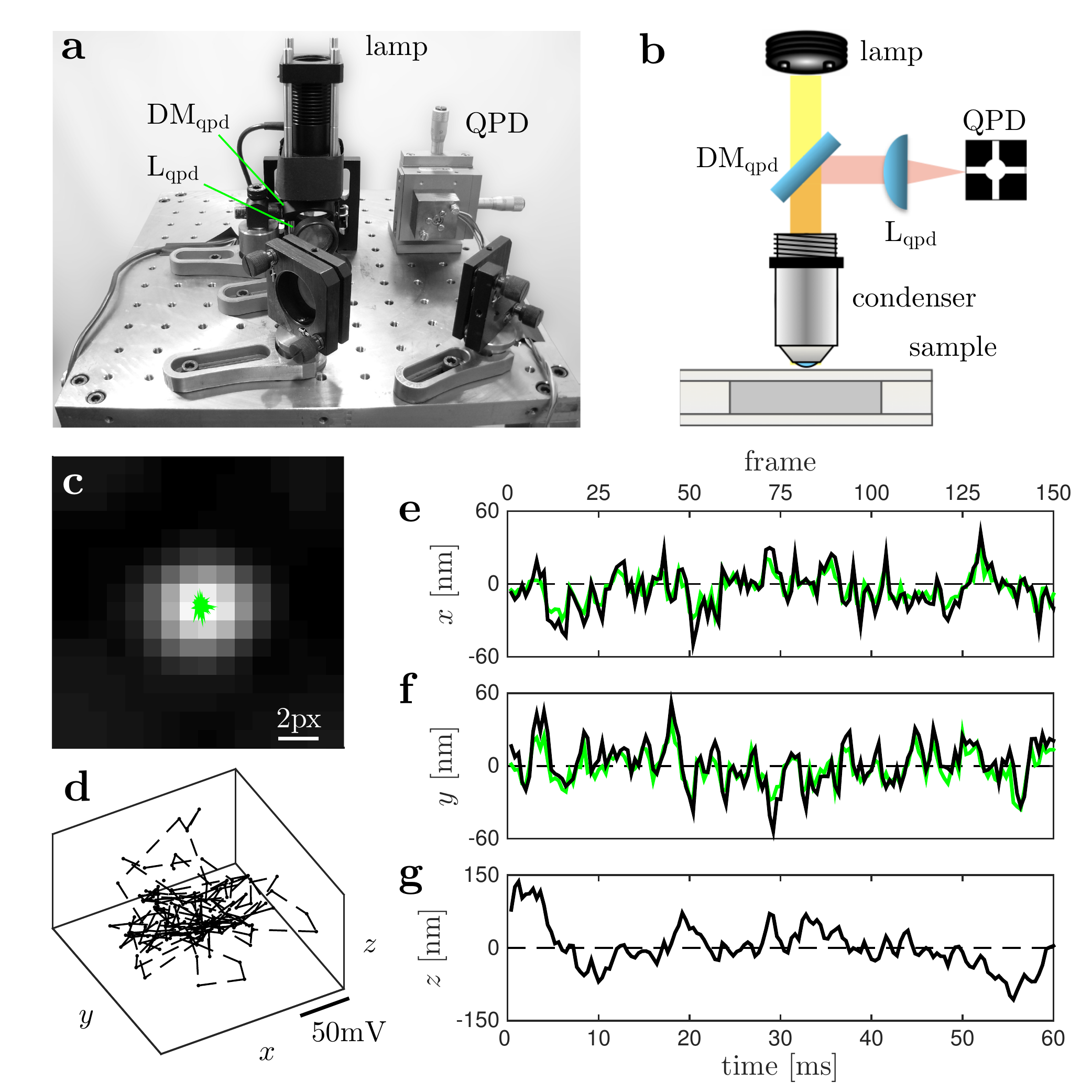}}
\caption{Photonic force microscope (PFM). (a) Photograph and (b) schematic of the detection apparatus that permits us to project the interference pattern at the back-focal plane (BFP) of the condenser lens onto a quadrant photodiode (QPD). The position of the particle can be monitored also using (c) digital video microscopy (DVM) or (d) the three interferometric signals received at the QPD. We obtain an excellent agreement between the positions measured with the two techniques as shown in panels (e) and (f), where the horizontal ($x$ and $y$) coordinates of the particle using DVM (green/gray solid lines) are overlapped to those obtained from interferometry (black solid lines). (g) The QPD permits to measure also the vertical ($z$) coordinate of the particle.}
\label{fig:qpd}
\end{figure}

\begin{enumerate}
\itemsep1em
\addtocounter{enumi}{24}

\item In order to use a QPD, the image plane of the condenser lens should be the same as the one of the objective, i.e., they should form a telescope. In order to achieve this, adjust the height of the condenser until the laser beam emerging from the condenser is perfectly collimated.

\item Position a dichroic mirror on the third level below the illumination LED (DM$_{\textrm{qpd}}$, Fig.~\ref{fig:qpd}a and \ref{fig:qpd}b).

\item Immediately after the dichroic mirror, place a $20\,{\rm cm}$ or $10\,{\rm cm}$ focal length lens
(L$_{\textrm{qpd}}$, Fig.~\ref{fig:qpd}a and \ref{fig:qpd}b) to project the image of the condenser BFP on the QPD. 

\item Mount the QPD on a three-axis translation stage for precise alignment. Depending on the space available on the third level, use mirrors to deflect the laser light towards the QPD (two mirrors in our case, Fig.~\ref{fig:qpd}a and \ref{fig:qpd}b). The QPD sensitive area should be as perpendicular to the laser beam as possible.

\item Proceed to connect the output of the QPD to the computer or to a digital oscilloscope to observe the output signals. 

\item Proceed to align the QPD. This involves first of all roughly centring the light collected by the condenser on the QPD and then proceeding to maximize the QPD sum signal while zeroing the differential QPD signals. This should be done without any trapped particle.

\item Pick a particle and observe the three signals. You should observe the thermal motion of the particles plus a small offset (Fig.~\ref{fig:qpd}d). Typically, the smaller the offsets the better the alignment.

\item Acquire the position signals of a particle as a function of time (Figs.~\ref{fig:qpd}e, \ref{fig:qpd}f and \ref{fig:qpd}g) and analyze them according to the procedures presented in Appendix~A in order to calibrate the optical trap. It is also possible to compare the trajectories obtained with the QPD to the ones obtained though digital video microscopy (Figs.~\ref{fig:qpd}e and \ref{fig:qpd}f). The agreement between the two results is an indication of the quality and stability of the set-up.

\end{enumerate}

\subsection{Holographic optical tweezers (HOT)}

\begin{figure}[ht]
\centerline{\includegraphics[width=0.7\columnwidth ]{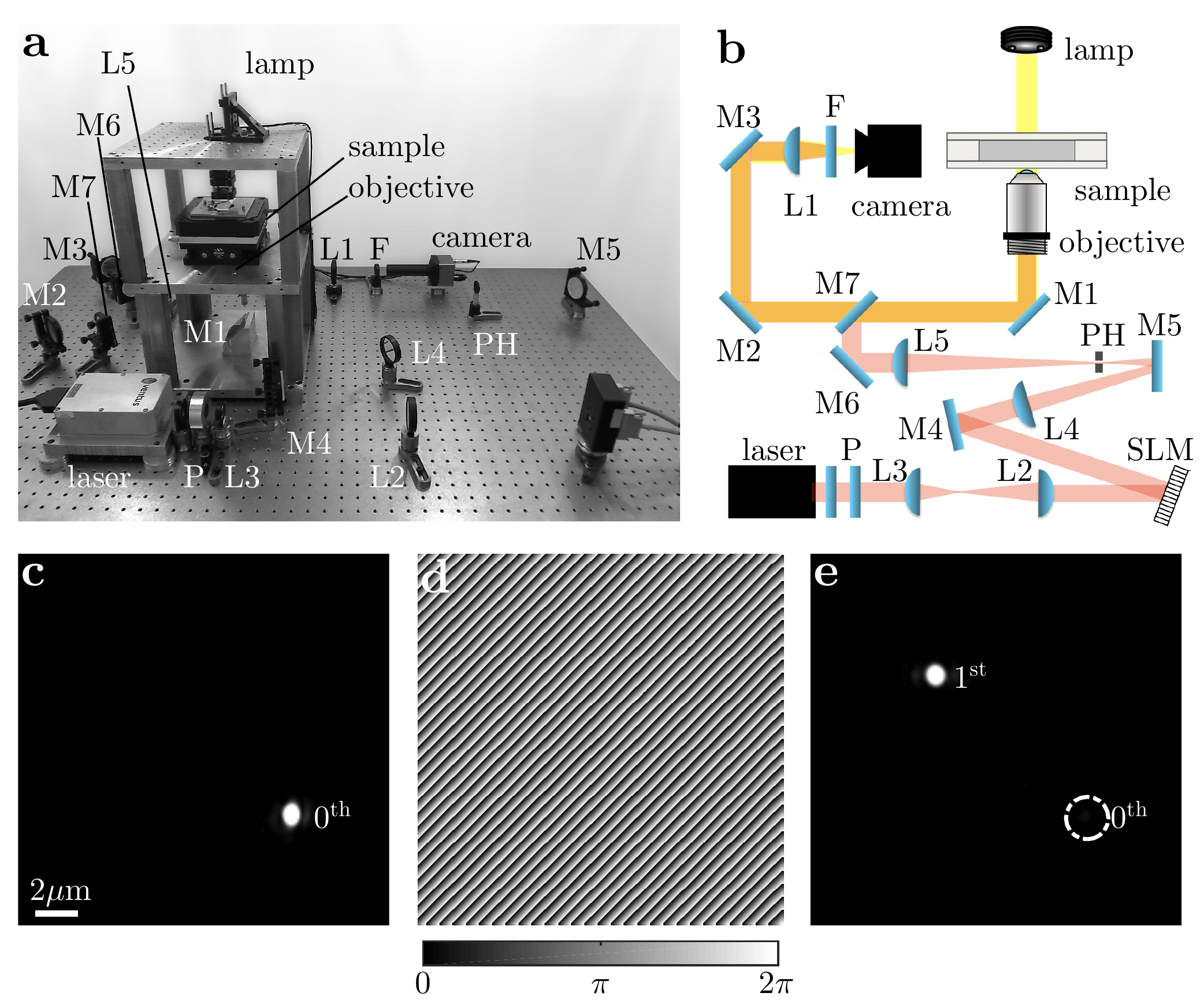}}
\caption{Holographic optical tweezers (HOT). (a) Photograph and (b) schematic of the experimental set-up. See Movie~2 for the complete building sequence. (c) In the absence of phase modulation on the spatial light modulator (SLM), the SLM works essentially as a mirror, reflecting almost the whole impinging light into a $0^{\rm th}$-order beam. When (d) a blazed grating phase profile is imposed on the SLM, (e) most of the beam light is deflected into a $1^{\rm st}$-order beam, while there is only some residual light in the $0^{\rm th}$-order beam (circled). The amount of light in the  $0^{\rm th}$-beam can be further reduced by the use of a pinhole (PH) placed at the beam waist between lenses L4 and L5.}
\label{fig:hot}
\end{figure}

\begin{figure}[ht]
\centerline{\includegraphics[width=0.7\columnwidth ]{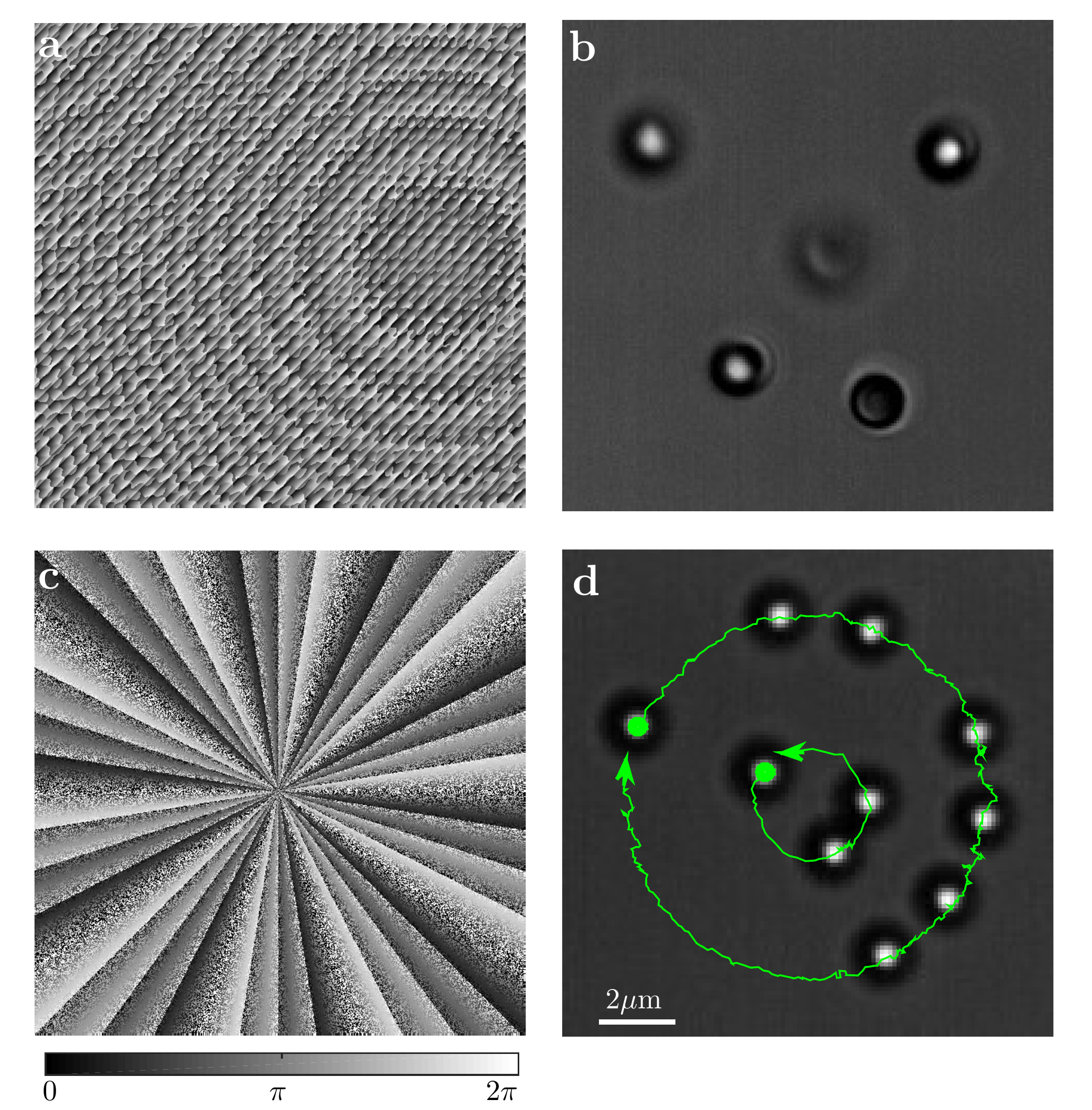}}
\caption{HOT at work. (a) Phase mask obtained by a random superposition of diffraction gratings and Fresnel lenses and (b) corresponding multiple optical traps placed at various lateral and vertical positions. (c) Phase mask to generate two Laguerre-Gaussian beams with topological charge of +10 (inner circle) and -40 (outer circle), and (d) corresponding optically trapped particles; in this case, due to the strong scattering forces, the beads cannot be trapped in three dimension but are pushed towards the top glass slide, where the particles rotate along two circles in opposite direction due to the transfer of orbital angular momentum (the solid lines represent the trajectory of two particles).}
\label{fig:mask}
\end{figure}

\begin{enumerate}
\itemsep1em
\addtocounter{enumi}{32}

\item In order to realize a HOT, we need to change the optical beam input train. In particular, we need to place the SLM in a plane conjugate to the back aperture of the objective using two lenses arranged in a $4f$-configuration. In order to fit this optical train within the optical table, it is necessary to carefully plan the laser beam path. In our case (Fig.~\ref{fig:hot}a and \ref{fig:hot}b), the laser beam is first reflected by the SLM, then by a series of mirrors (M4, M5, M6, M7, M1), and finally it reaches the objective. The total distance between the SLM and the objective back aperture is $3000\,{\rm mm}$, so that we can realize a $4f$-configuration using two lenses with $f = 750\,{\rm mm}$ (L4 and L5, step~41).

\item Change the illumination system in order to increase the light intensity. In particular, place the LED light close to the sample and use a couple of aspherical lenses to collimate the LED light.

\item Repeat steps~14-15 to place the laser. Notice that to create multiple traps more laser power is required. Therefore we used a different DPSS laser with a maximum power of $3\,{\rm W}$ (Laser Quantum, Ventus 1064).

\item Expand the optical beam with a two-lens telescope (L2 and L3, Fig.~\ref{fig:hot}a and \ref{fig:hot}b) as explained in step~18, so that it fills the SLM active area. The laser spot should be perfectly centered on the SLM active area, lest the efficiency and the wavefront shape will be affected.

\item Rotate the SLM by a small (less than \ang{10}) angle with respect to the incidence direction of the laser beam. Larger angles would result in a significant decrease of the SLM efficiency.

\item Display a blazed diffraction grating on the SLM. The SLM will split the incoming beam into several diffraction orders. Usually the strongest are the $0^{\rm th}$-order beam, i.e., the beam that is simply reflected by the active area, and the $1^{\rm st}$-order diffracted beam, which is the one that will be used to generate the optical traps. The intensity of the $1^{\rm st}$-order beam varies according to the efficiency of the SLM.

\item Use large optics (at least $2\,{\rm inch}$ diameter) in the optical train in order to collect as much as possible of the light modulated into the $1^{\rm st}$-order beam.

\item Deflect the laser beam towards the objective back aperture. Align the laser beam so that it goes straight through the objective holder using mirrors M6 and M7 (Fig.~\ref{fig:hot}a and \ref{fig:hot}b), as done in step~17.

\item Add lenses L4 and L5 ($f = 750\,{\rm mm}$, Fig.~\ref{fig:hot}a and \ref{fig:hot}b) to realize the $4f$-configuration. Lens L5 must be placed exactly at a distance from the back  aperture of the objective lens equal to $f$. While aligning L5 check that the position of the laser beam on the alignment tools does not change. Then, place L4 at a distance of $1500\,{\rm mm}$ from the first lens and of $750\,{\rm mm}$ from the SLM surface. Again, check that the position of the laser spot does not change after both lenses have been positioned.

\item The middle point between L4 and L5 is conjugated to the objective FFP and, thus, it is possible to place a pinhole at this position to select only the $1^{\rm st}$-order beam and cut out the $0^{\rm th}$-order beam and the higher diffraction orders.
 
\item Repeat steps~20-24 to prepare the sample cell and observe the particles.

\item Observe how the optical trap can be moved in the lateral direction by using gratings with different orientations and pitches, and in the axial direction by using a Fresnel lens (Appendix~B). Note also how with a combination of gratings and Fresnel lenses it is possible to displace a trap in three-dimensions. Generate multiple optical traps and trap various particles, as in Figs.~\ref{fig:mask}a and \ref{fig:mask}b. Generate Laguerre-Gaussian beams and see the transfer of orbital angular momentum (OAM), as in Figs.~\ref{fig:mask}c and \ref{fig:mask}d.

\end{enumerate}

\subsection{Speckle optical tweezers}

\begin{figure}[ht]
\centerline{\includegraphics[width=0.7\columnwidth ]{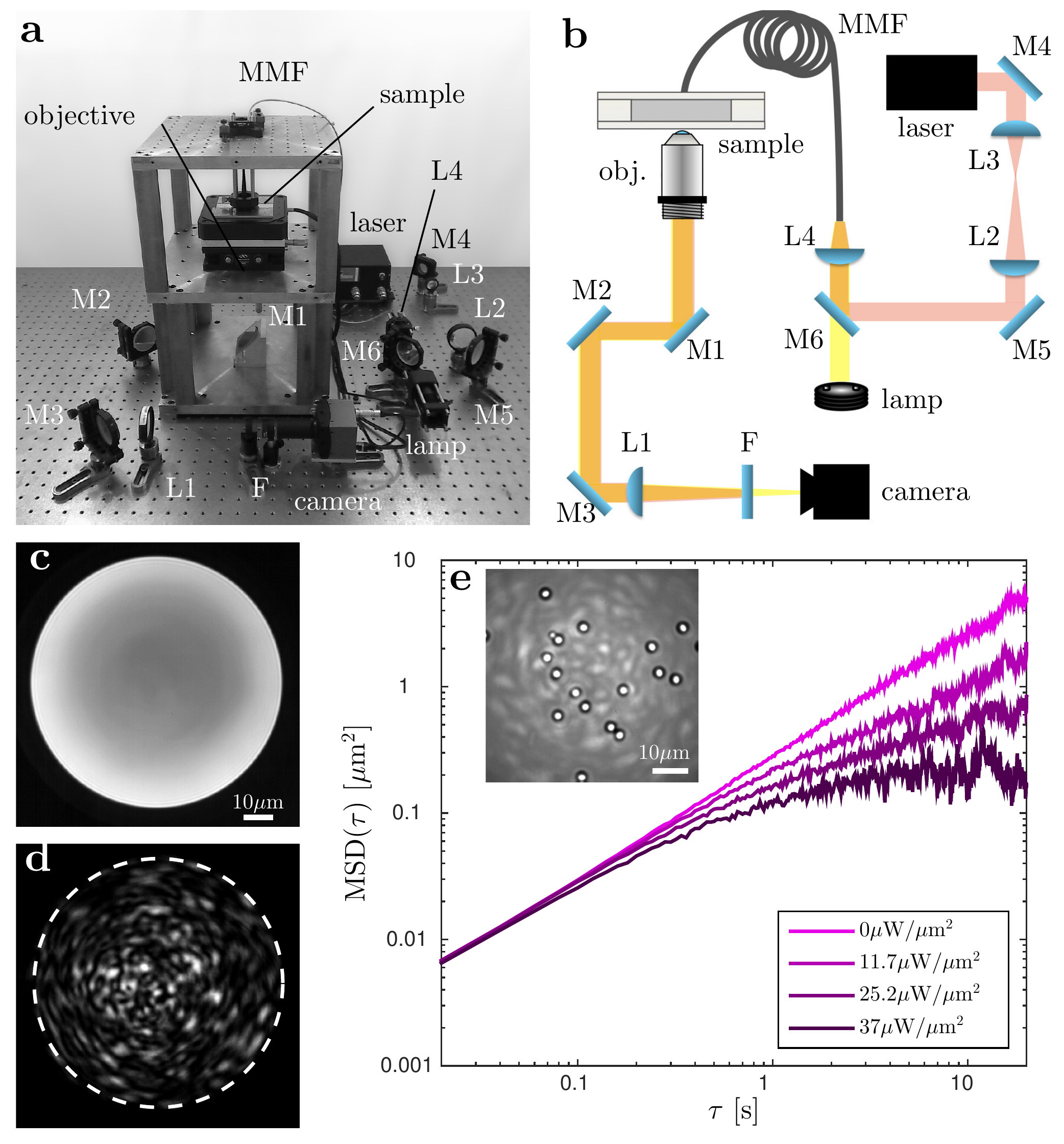}}
\caption{Speckle optical tweezers. (a) Photograph and (b) schematic of the experimental set-up. See Movie~3 for the complete building sequence. (c) Illumination light and (d) speckle light field output from the optical fiber. (e) Mean square displacement (MSD) of particles trapped at various laser powers; a transition from trapping at high laser powers to subdiffusion at low laser powers can be observed.}
\label{fig:speckle}
\end{figure}

\begin{enumerate}
\itemsep1em
\addtocounter{enumi}{44}

\item Repeat steps~14-15 to place the laser.

\item Mount a telescope to increase the size of the laser beam waist (L2 and L3, Fig.~\ref{fig:speckle}a and \ref{fig:speckle}b).

\item It is necessary to overlap the illumination light on the laser beam to have them together at the fiber output. In order to do this use mirrors  (M5 and M6, Fig.~\ref{fig:speckle}a and \ref{fig:speckle}b) to steer the laser beam towards the entrance port of the fiber. One mirror (in our case, M6) must be a dichroic mirror to allow the transmission of the white light of the illumination LED.

\item Align the laser beam into the fiber (Thorlabs M15L01, fiber core $105\,{\rm \mu m}$ in diameter) input using a lens (L4, Fig.~\ref{fig:speckle}a and \ref{fig:speckle}b). To maximize the coupling and to generate a fully developed speckle pattern, the focused beam should match the numerical aperture of the fiber. Since in our case the fiber has a numerical aperture of 0.20, we used a $20\,{\rm mm}$ focal length lens with $12.7\,{\rm mm}$ and, in order to match the numerical aperture of the fiber, the beam waist was set to $4\,{\rm mm}$. We suggest to built a cage system to firmly hold the focusing lens and the fiber end. The fiber end, or equivalently the focusing lens, can be mounted on a translation stage (Thorlabs, SMZ1) to easily align the beam. Using a power-meter, maximize the power at the output end of the fiber.

\item Place the illumination  LED behind the dichroic mirror (M6, Fig.~\ref{fig:speckle}a and \ref{fig:speckle}b) and observe the white light at the end of the fiber with the laser light off. Optimize the transmission of the illumination LED using the CCD camera. This can be easily done by moving the whole LED mount behind the dichroic mirror.

\item Using an IR card, you should now be able to observe a diverging speckle pattern from the output of the fiber, as shown in Fig.~\ref{fig:speckle}d.

\item Build a cage system to host the output fiber adapter. It should be mounted vertically upside down just above the sample cell. Mount it on a precision three-axis translation stage to align the fiber output face along the optical axis of the observation objective (Fig.~\ref{fig:speckle}a and \ref{fig:speckle}b) and to control the size of the average speckle grain by precisely translating the fiber vertically. 

\item Place this system on the third level of the microscope (Fig.~\ref{fig:speckle}a).

\item In order to be able to observe the whole area corresponding to the fiber core ($105\,{\rm \mu m}$ in diameter), use an imaging objective with a relatively low magnification (we used a $40\times$ objective).

\item The sample cell should be realized in a slight different way the one followed in step~23: since the output fiber end must be very close to the sample  ($< 300\,{\rm \mu m}$), the $1\,{\rm mm}$ glass slide needs to be replaced by a much thinner coverslip. Place the sample cell on the piezoelectric translation stage and move the objective until you observe the particles on the bottom of the sample cell. 

\item Change the height of the output fiber end until you can see its clear image (Fig.~\ref{fig:speckle}c). Then move it back by a very small amount. You should observe, using the CCD camera, a blurred bright disk.  

\item Repeat step~20 to filter out the laser light. Note that in this case the laser light collected by the objective is sent towards the CCD. It could be necessary to use multiple filters to eliminate all the laser light.

\item Set the laser output power at a moderate level. Remove the short-pass filter from in front of the CCD camera. The speckle pattern will be now visible inside the bright disk imaged by the camera, as shown in Fig.~\ref{fig:speckle}d.

\item At this point you should observe that some of the free diffusing particles on the bottom of the sample cell (inset in Fig.~\ref{fig:speckle}e) are attracted towards the high intensity maxima of the speckle pattern. Increasing the laser power will increase the number of hot spots with enough power to actually trap particles.

\item Measure the trajectories of the particles using digital video microscopy.

\item Observe how increasing the laser power the particles are more and more confined in the speckle grains. Use, e.g., the mean square displacement analysis (see Appendix B) to quantify how much the particles are trapped, as shown in Fig.~\ref{fig:speckle}e.

\item Moving the piezoelectric stage, observe how the speckle pattern captures particles as they approach the high intensity hot spots.

\item Prepare a binary or ternary mixture of particles of different size and/or different material. Observe how the speckle optical tweezers is able to selectively capture different particles as the piezoelectric stage is moved.

\end{enumerate}

\section{Conclusions}

In this article we presented a detailed step-by-step guide for the construction of advanced optical manipulation techniques based on the construction of a home-made inverted optical microscope. Readers with some basic experience in experimental optics should be able to follow  the explained procedure and build a single-beam OT, a HOT and, finally, a speckle OT. Readers will also be able to easily upgrade the designs we propose and realize in this article in order to fit their experimental needs. 

Giovanni Volpe was partially supported by Marie Curie Career Integration Grant (MC-CIG) under Grant PCIG11 GA-2012-321726 and a T\"UBA-GEBIP Young Researcher Award.

\appendix

\section{Calibration of the single-beam optical trap}

\begin{figure}[ht]
\centerline{\includegraphics[width=0.7\columnwidth ]{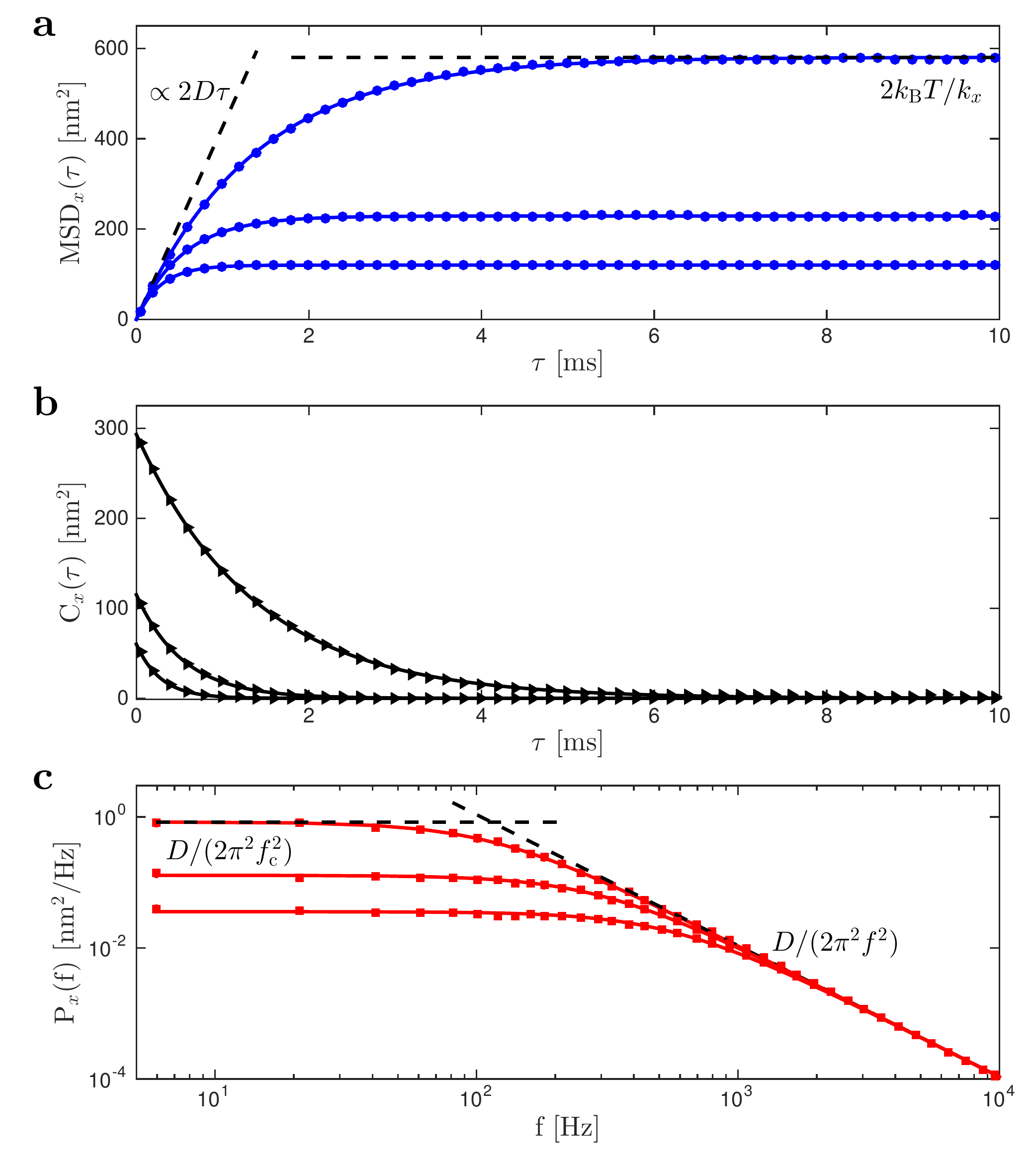}}
\caption{Calibration of an optical trap using the (a) mean square displacement (MSD), (b) autocorrelation function (ACF) and (c) power spectral density (PSD) methods. The symbol correspond to the MSD, ACF and PSD experimentally obtained from a $10\,{\rm s}$ trajectory of a $1\,{\rm \mu m}$ diameter polystyrene particle trapped by a  single-beam optical tweezers. The solid lines correspond to the best fit to the theoretical MSD (Eq.~(\ref{eq:msd})), ACF (Eq.~(\ref{eq:acf})) and PSD (Eq.~(\ref{eq:psd})).}
\label{fig:analysis}
\end{figure}

\begin{figure}[ht]
\centerline{\includegraphics[width=0.7\columnwidth ]{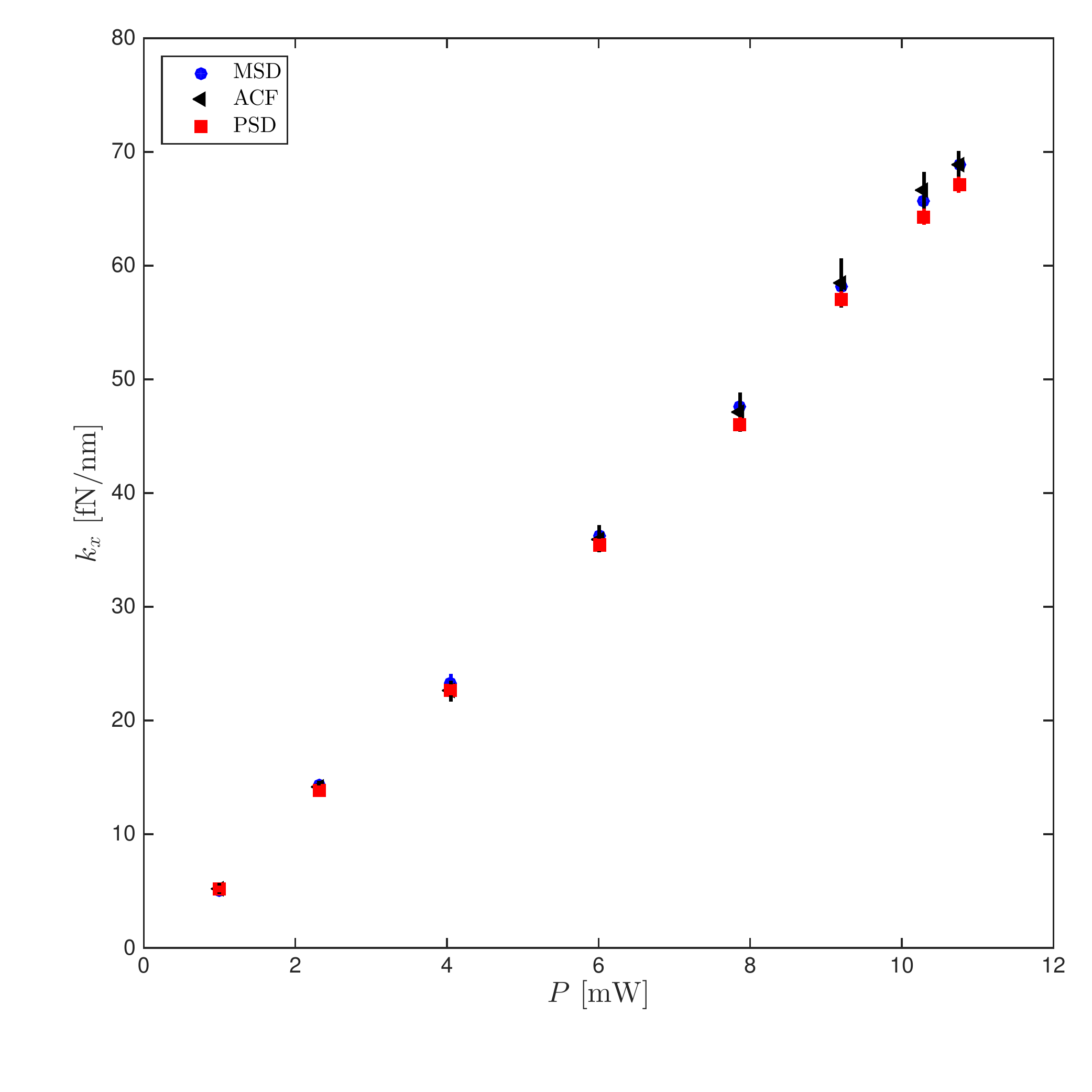}}
\caption{Linearity of the trap stiffness $k_x$ as a function of the laser power at the trap $P$. The calibrations obtained with MSD, ACF and PSD show an excellent agreement.}
\label{fig:otcalib}
\end{figure}

Once the trajectory of a Brownian particle has been measured either by digital video microscopy or by interferometry, it is possible to use it in order to study quantitatively the optical potential. In calibrating an optical trap, the main objective is to determine the trap stiffness $k_x$ along each direction, in this case, e.g., along $x$ (one of the horizontal directions). A common secondary objective is to determine the conversion factor $\beta_x$ from measurement units (e.g., pixels or volts) to physical units of  length (meters). The knowledge of these two parameters allows to measure an external force applied to the trapped particle by measuring only  the bead displacement from its equilibrium position. Several techniques can be employed for this task \cite{FlorinAPA98,NeumanRSI04}. The easiest and fastest is based on the equipartition theorem which, for a harmonic potential, predicts an energy equal to $\frac{1}{2}k_BT$ for each each degree of freedom. The trajectory along one coordinate has an energy of $\frac{1}{2}k_x \langle x^2 \rangle$, thus from a direct comparison it is possible to determine the trap stiffness. A more detailed determination uses the Boltzmann statistics, where the optical potential well is obtained from the  probability distribution associated to a particle's trajectory.  However, the most common and reliable methods employ the mean square displacement (MSD), the autocorrelation function (ACF) and the power spectral distribution (PSD). We will briefly discuss them in this Appendix.

The MSD provides a measure of the displacement of a particle from its initial position. The MSD as function of the lag time $\tau$ is given by ${\rm MSD}_x(\tau)=\langle \left(x(t+\tau)-x(t)\right)^2 \rangle$. For free particles diffusing in a viscous medium, the MSD grows linearly with time, so that ${\rm MSD}_x (\tau)=2D\tau$, where $D$ is the diffusion coefficient. A particle held in an optical tweezers is continuously pulled back by the restoring force, therefore the MSD cannot grow indefinitely but reaches a plateau, so that
\begin{equation}\label{eq:msd}
{\rm MSD}_x (\tau)=\frac{2 k_BT}{k}\left[ 1-e^{-\frac{k \tau}{\gamma} }\right] ,
\end{equation}
from which we can see that the particle diffuses inside the optical potential well at short times, i.e., until it starts feeling the trap restoring force. An OT can be calibrated by fitting the experimentally obtained MSD to Eq.~(\ref{eq:msd}), as shown in Fig.~\ref{fig:analysis}a. The plateau at long times depends on $k$, while the linear growth at very short times depends, at a first approximation \cite{LukicPRL05,PescePRE14}, on $D$.

The ACF provides a measure of how fast a particle falls into the optical trap. The ACF on an optically trapped particle is 
\begin{equation}\label{eq:acf}
\mathcal{C}_x(\tau)=\frac{k_BT}{k_x}e^{-\frac{k_x \tau}{\gamma}},
\end{equation}
where $\gamma$ is the friction coefficient. The trap stiffness can be straightforwardly determined from a fit procedure of the ACF calculated from the experimental trajectories as shown in Fig.~\ref{fig:analysis}b. As the trap stiffness increases the exponential decay of the ACF becomes faster. We note that also cross-correlation functions, i.e., correlation between the motion of a particle in two different directions, can be used to calibrate an optical trap as they can provide information about the presence of cross-talk \cite{SorensenRSI04} and about the presence of torques \cite{VolpePRL06}.

The PSD is the Fourier transform of the ACF and, for an optically trapped particle, it is
\begin{equation}\label{eq:psd}
P_x(f)=\frac{k_BT}{2\pi^2\gamma}\frac{1}{f^2+f_{\rm c,x}^2},
\end{equation}
where $f_{\rm c}=2\pi\gamma k_x$ is the corner frequency, which represents the frequency beyond that the motion is purely diffusive. The calibration of the optical trap can again be achieved by fitting the experimentally obtained PSD to the theoretical one, as shown in Fig.~\ref{fig:analysis}c. As the optical trap stiffness increases, $f_{\rm c}$ increases. One advantage of the PSD analysis is that it permits one to clearly identify various sources of noise, e.g., electrical interferences and mechanical drifts \cite{SorensenRSI04}. In particular, the flat plateau at low frequencies of the PSD in Fig.\ref{fig:analysis}c  is a clear sign of the high mechanical stability of the system.

The data reported in Fig.~\ref{fig:otcalib} show the linear dependence of the trap stiffness determined using the methods described above on the trapping laser power. The data are obtained using the single-beam OT with interferometric detection with the QPD.  The agreement between the different techniques is excellent and demonstrate the quality of our set-up.

\section{Generation of phase masks for HOT}

The working principle of  HOT is based on a DOE whose task is to shape the profile of an  incoming optical beam. The easiest way to realize such an element  is to use an SLM. A SLM is a transmissive or reflective device that is used to spatially modulate the amplitude and phase of an optical wavefront in two dimensions. SLMs can be programmed to produce light beams with various optical wavefronts and can, therefore, be used in place of conventional optical elements, such as lenses, mirror and gratings. Furthermore, they can be used to create optical vortex and higher order Bessel beams, which feature an azimuthal phase variation. To accomplish these tasks, it is in principle necessary to display on the SLM  a hologram obtained taking the inverse Fourier  transform of the desired intensity distribution in the front focal plane. However, the simultaneous modulation of phase and amplitude are technically difficult and, furthermore, the amplitude modulation would remove power from the beam, reducing the modulation efficiency. Therefore, in practice, it is more convenient to employ a phase-only modulation hologram.

The SLM is typically placed in a plane conjugated to the objective BFP so that the light distribution generated on the FFP is essentially the Fourier transform of the phase mask imposed on the SLM (assuming an incoming beam with a flat wavefront). Therefore, imposing a phase mask corresponding tot a blazed diffraction grating will change the angle of the beam emerging from the SLM and will displace the focus at the image plane; the angle of the grating determines the direction of the displacement, while the number of fringes determines the distance.  To change the position of the focus along the axial direction, it is necessary to change the curvature of the wavefront and this is accomplished applying to the SLM the pattern of a Fresnel lens. A combination of a diffraction grating and a Fresnel lens produces a shift of the trap in three dimensions, more precisely, the corresponding phase mask is given by:
\begin{equation}\label{eq:hot:single}
\phi^{\rm S}(x,y) = 
\underbrace{\frac{2\pi}{\lambda_0 f} \left[ x x_{{\rm to},1} + y y_{{\rm to},1} \right]}_{\mbox{Diffraction grating}}
+
\underbrace{\frac{\pi z_{{\rm to},1}}{\lambda_0 f^2} \left[ x^2 + y^2  \right]}_{\mbox{Fresnel lens}},
\end{equation}
where $x$ and $y$ are the cartesian coordinates on the SLM, $\lambda_0$ is the vacuum wavelength, $f$ is the focal distance of the lens and $[x_{{\rm to},1},y_{{\rm to},1},z_{{\rm to},1}]$ is the desired position of the trap in the image plane. Since the generation of this holographic mask is very fast, this is a very effective way to generate a single optical trap that can be moved in three dimensions in real time.

In order to generate multiple ($N$) traps, one of the fastest, but also least efficient, algorithm is the random mask encoding algorithm \cite{MontesOE06}. For every pixel of the SLM, a phase shift is determined as if the hologram was made to generate only one of the $N$ traps. This technique is very fast and permits one to achieve a good uniformity amongst traps. Nevertheless, the overall efficiency can be very low when $N$ is large, because on average the numbers of pixels that interfere constructively to generate each trap decreases as $1/N$.

A straightforward algorithm, which achieves a better efficiency than the random mask encoding at a just 
slightly higher computational cost, is the superposition of gratings and lenses algorithm \cite{DiLeonardoOE11}. In this algorithm, the phase of each pixel is chosen equal to the argument of the complex sum of 
single-trap holograms. This algorithm has typically a good efficiency at the cost of poor uniformity. In fact, a 
typical problem that arises using this algorithm is that, when highly symmetrical trap geometries are sought, 
such as an array of traps, a consistent part of the energy is diverted to unwanted ghost traps.

To overcome these difficulties it is necessary to use more sophisticated algorithms to compute the phase 
patterns. The most used are the Gerchberg-Saxton \cite{GerchbergO72} and the adaptive-additive algorithms 
\cite{DufresneRSI01}, that are able to realize a hologram that will lead to a very close approximation to any desired 
intensity distribution. The Gerchberg-Saxton algorithm is an iterative algorithm that permits one to find a 
phase distribution that turns a given input intensity distribution arriving at the hologram plane into a desired 
intensity distribution in the trapping plane by propagating the complex amplitude back and forth between 
these two planes and replacing at each step the intensity on the trapping plane with the target intensity and 
that on the SLM plane with the laser's actual intensity profile. It typically converges after a few tens of 
iterations, it is ideally suited to deal with continuous intensity distributions. The adaptive-additive algorithm 
starts with an arbitrary guess of the phase profile and an initial input wavefront,  the Fourier transform of this 
wavefront is the starting estimate for the output electric field; the resulting error in the focal plane is reducing 
by mixing a proportion of the desired amplitude into the field in the focal plane; inverse-Fourier transforming 
the resulting focal field yields the corresponding field in the input plane; the amplitude in the input plane is 
replaced with the actual amplitude of the laser profile; finally, the algorithm is iterated. The main advantage 
of this algorithm in comparison to the Gerchberg-Saxton algorithm is that permits one to achieve a better 
uniformity over all the traps in the array.

SLM can also be used to create non-Gaussian beams such as Hermite-Gaussian beams, Laguerre-Gaussian beams, non-diffracting beams  and continuous intensity distributions. In particular, the phase mask to generate  Laguerre-Gaussian beams is given by:
\begin{equation}\label{eq:hot:lg}
\phi^{\rm LG} (\varphi,\rho) = i l \varphi,
\end{equation}
where $\varphi$ and $\rho$ are the polar coordinated of a system of reference centered on the middle of the SLM and $l$ is the order to the Laguerre-Gaussian beam.


\end{document}